\documentclass[aps,prl,preprint,superscriptaddress,showpacs]{revtex4}

\usepackage{graphicx}

\begin{document}


\title{Probing shell structure and shape changes in
neutron-rich sulfur isotopes through transient-field $g$-factor
measurements on fast radioactive beams of $^{38}$S and $^{40}$S.}




\author{A.D.~Davies}
\affiliation{National Superconducting Cyclotron Laboratory, Michigan
State University, East Lansing, MI 48824 USA}
\affiliation{Department of Physics and Astronomy, Michigan State
University, East Lansing, MI 48824 USA}
\author{A.E.~Stuchbery}
\affiliation{Department of Nuclear Physics, The Australian National
University, Canberra, ACT 0200, Australia}
\author{P.F.~Mantica}
\affiliation{National Superconducting Cyclotron Laboratory, Michigan
State University, East Lansing, MI 48824 USA}
\affiliation{Department of Chemistry, Michigan State University,
East Lansing, MI 48824 USA}
\author{P.M.~Davidson}
\affiliation{Department of Nuclear Physics, The Australian National
University, Canberra, ACT 0200, Australia}
\author{A.N.~Wilson}
\affiliation{Department of Nuclear Physics, The Australian National
University, Canberra, ACT 0200, Australia} \affiliation{Department
of Physics, The Australian National University, Canberra, ACT 0200,
Australia}
\author{A.~Becerril}
\affiliation{National Superconducting Cyclotron Laboratory, Michigan
State University, East Lansing, MI 48824 USA}
\affiliation{Department of Physics and Astronomy, Michigan State
University, East Lansing, MI 48824 USA}
\author{B.A.~Brown}
\affiliation{National Superconducting Cyclotron Laboratory, Michigan
State University, East Lansing, MI 48824 USA}
\affiliation{Department of Physics and Astronomy, Michigan State
University, East Lansing, MI 48824 USA}
\author{C.M.~Campbell}
\affiliation{National Superconducting Cyclotron Laboratory, Michigan
State University, East Lansing, MI 48824 USA}
\affiliation{Department of Physics and Astronomy, Michigan State
University, East Lansing, MI 48824 USA}
\author{J.M.~Cook}
\affiliation{National Superconducting Cyclotron Laboratory, Michigan
State University, East Lansing, MI 48824 USA}
\affiliation{Department of Physics and Astronomy, Michigan State
University, East Lansing, MI 48824 USA}
\author{D.C.~Dinca}
\affiliation{National Superconducting Cyclotron Laboratory, Michigan
State University, East Lansing, MI 48824 USA}
\affiliation{Department of Physics and Astronomy, Michigan State
University, East Lansing, MI 48824 USA}
\author{A.~Gade}
\affiliation{National Superconducting Cyclotron Laboratory, Michigan
State University, East Lansing, MI 48824 USA}
\author{S.N.~Liddick}
\affiliation{National Superconducting Cyclotron Laboratory, Michigan
State University, East Lansing, MI 48824 USA}
\affiliation{Department of Chemistry, Michigan State University,
East Lansing, MI 48824 USA}
\author{T.J.~Mertzimekis}
\affiliation{National Superconducting Cyclotron Laboratory, Michigan
State University, East Lansing, MI 48824 USA}
\author{W.F.~Mueller}
\affiliation{National Superconducting Cyclotron Laboratory, Michigan
State University, East Lansing, MI 48824 USA}
\author{J.R.~Terry}
\affiliation{National Superconducting Cyclotron Laboratory, Michigan
State University, East Lansing, MI 48824 USA}
\affiliation{Department of Physics and Astronomy, Michigan State
University, East Lansing, MI 48824 USA}
\author{B.E.~Tomlin}
\affiliation{National Superconducting Cyclotron Laboratory, Michigan
State University, East Lansing, MI 48824 USA}
\affiliation{Department of Chemistry, Michigan State University,
East Lansing, MI 48824 USA}
\author{K.~Yoneda}
\affiliation{National Superconducting Cyclotron Laboratory, Michigan
State University, East Lansing, MI 48824 USA}
\author{H.~Zwahlen}
\affiliation{National Superconducting Cyclotron Laboratory, Michigan
State University, East Lansing, MI 48824 USA}
\affiliation{Department of Physics and Astronomy, Michigan State
University, East Lansing, MI 48824 USA}


\date{\today}

\begin{abstract}
The shell structure underlying shape changes in neutron-rich nuclei
near $N=28$ has been investigated by a novel application of the
transient field technique to measure the first-excited state
$g$~factors in $^{38}$S and $^{40}$S produced as fast radioactive
beams. There is a fine balance between proton and neutron
contributions to the magnetic moments in both nuclei. The $g$~factor
of deformed $^{40}$S does not resemble that of a conventional
collective nucleus because spin contributions are more important
than usual.
\end{abstract}



\maketitle



The fundamental question of how major shell closures, or magic
numbers, change in neutron-rich nuclei remains unresolved. At
present there is conflicting evidence concerning the $N=28$ shell
gap in nuclei approaching the neutron dripline. From the measured
$\beta$-decay halflife it has been suggested that
$^{42}_{14}$Si$_{28}$ is strongly deformed, implying a quenching of
the $N=28$ gap \cite{gre04}, whereas knockout reactions on $^{42}$Si
give evidence for a nearly spherical shape \cite{fri05}.
Low-excitation level structures and $B(E2)$ values imply that the
nearby even sulfur isotopes between $N=20$ and $N=28$ undergo a
transition from spherical at $^{36}_{16}$S$_{20}$, to prolate
deformed in $^{40}_{16}$S$_{24}$ and $^{42}_{16}$S$_{26}$, and that
the $N=28$ nucleus $^{44}_{16}$S$_{28}$ appears to exhibit
collectivity of a vibrational
character~\cite{Sche1996,Glasm1997,Wing2001,Sohl2002}. However the
evolution of deformation in these nuclei has underlying causes that
remain unclear. Some have argued that a weakening of the $N=28$
shell gap is important \cite{Sohl2002}, while others have argued
that the effect of adding \textit{neutrons} to the $f_{7/2}$ orbit
is primarily to reduce the \textit{proton} $s_{1/2}$-$d_{3/2}$ gap
and that a weakening of the $N=28$ shell gap is not needed to
explain the observed collectivity near $^{44}$S \cite{Cott1998}.
There have been several theoretical studies discussing the erosion
of the $N=28$ shell closure and the onset of deformation (e.g.
Refs.~\cite{rod02,cau04} and references therein).

To resolve questions on the nature and origins of deformation near
$N=28$, we have used a novel technique to measure the $g$ factors of
the 2$_1^+$ states in $^{38}_{16}$S$_{22}$ and $^{40}_{16}$S$_{24}$.
The $g$~factor, or gyromagnetic ratio, is the magnetic moment
divided by the angular momentum. The existence of deformation in
nuclei has long been associated with strong interactions between a
significant number of valence protons and neutrons, particularly in
nuclei near the middle of a major shell. Without exception the
deformed nuclei studied to date have $g$~factors near the
hydrodynamical limit, $Z/A$, reflecting the strong coupling between
protons and neutrons, and a magnetic moment dominated by the orbital
motion of the proton charge with small contributions from the
intrinsic magnetic moments of either the protons or the neutrons.
In transitional regions $g$~factors have considerable sensitivity to
the proton and/or neutron contributions to the state wavefunctions,
particularly if the intrinsic spin moments of the nucleons come to
the fore.

The present Letter presents the first application of a high-velocity
transient-field (HVTF) technique \cite{Stuc2004a} to measure the
$g$~factors of excited states of fast radioactive beams.
In brief, intermediate-energy Coulomb excitation
\cite{Glasm1998} is used to excite and align the nuclear states of
interest. The nucleus is then subjected to the transient field in a
higher velocity regime than has been used previously for moment
measurements, which causes the nuclear spin to precess. Finally, the
nuclear precession angle, to which the $g$~factor is proportional,
is observed via the perturbed $\gamma$-ray angular correlation
measured using a multi-detector array. It is important to note that
the transient-field technique has sensitivity to the {\em sign} of
the $g$~factor, which in itself can be a distinguishing
characteristic of the proton/neutron contributions to the state
under study, since the signs of the spin contributions to the proton
and the neutron $g$~factors are opposite.


The transient field (TF) is a velocity-dependent magnetic hyperfine
interaction experienced by the nucleus of a swift ion as it
traverses a magnetized ferromagnetic material \cite{Kol1980,spe02}.
For light ions ($Z \leq 16$) traversing iron and gadolinium hosts at
high velocity, the dependence of the TF strength on the ion
velocity, $v$, and atomic number, $Z$, can be parametrized
\cite{Stuc2004a,Stuc2005a} as
\begin{equation}
B_{\rm tf}(v,Z) = A Z^P (v/Zv_0)^2 {\rm e}^{-
\frac{1}{2}(v/Zv_0)^4}, \protect \label{eq:aes-param}
\end{equation}
where $v_0 = c/137$ is the Bohr velocity. A fit to data for iron
hosts yielded $A=1.82(5)$~T with $P=3$ \cite{Stuc2004a}. The maximum
TF strength is reached when the ion velocity matches the $K$-shell
electron velocity, $v = Zv_0$. Since the transient field arises from
polarized electrons carried by the moving ion its strength falls off
as the ion velocity exceeds $Zv_0$ and becomes fully stripped; a
transient-field interaction will not occur for fast radioactive
beams with energies near 100 MeV/nucleon until most of that energy
is removed.


The experiment was conducted at the Coupled Cyclotron Facility of
the National Superconducting Cyclotron Laboratory at Michigan State
University.  Secondary beams of $^{38}$S and $^{40}$S were produced
from 140 MeV/nucleon primary beams directed onto a $\sim 1$~g/cm$^2$
$^9$Be fragmentation target at the entrance of the A1900 fragment
separator \cite{Morr2003}. An acrylic wedge degrader 971 mg/cm$^2$
thick and a 0.5\% momentum slit at the dispersive image of the A1900
were employed.  The wedge degrader allowed the production of highly
pure beams and also reduced the secondary beam energy to $\sim
40$~MeV/nucleon. Further details of the radioactive beams are given
in Table~\ref{tab:beams}. The $^{38}$S ($^{40}$S) measurement ran
for 81 (68) hours.

\begin{table}[htb]
\caption{Production and properties of radioactive beams.}
\begin{ruledtabular}
\begin{tabular}{cccccc}
\multicolumn{2}{c}{Primary beam} & \multicolumn{4}{c}{Secondary
beam}  \\ \cline{1-2} \cline{3-6}
Ion  & Intensity & Ion & $E$ & Intensity & Purity \\
 & (pnA) &  & (MeV) & (pps) & (\%)  \\ \hline
$^{40}$Ar & 25 & $^{38}$S & 1547.5 & $ 2 \cdot 10^5$ & $>99$ \\
$^{48}$Ca & 15 & $^{40}$S & 1582.5 & $ 2 \cdot 10^4$ & $>95$ \\
\end{tabular}
\end{ruledtabular}
\protect \label{tab:beams}
\end{table}

\begin{figure}[tbp]
\resizebox{86mm}{!}{\includegraphics{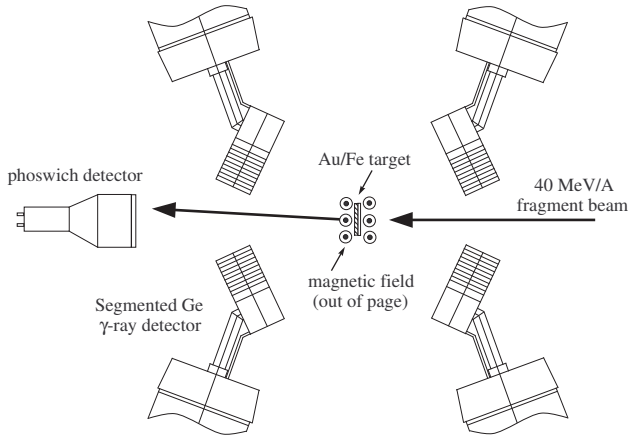}} \caption{Schematic
view of the experimental arrangement from above. Only the four SeGA
detectors perpendicular to the magnetic field axis are shown.  There
are ten other SeGA detectors (see text) which are not shown for
clarity.} \label{fig:schematic}
\end{figure}

Figure \ref{fig:schematic} shows the experimental arrangement. The
radioactive beams were delivered onto a target which consisted of a
355 mg/cm$^2$ Au layer backed by a 110 mg/cm$^2$ Fe layer of
dimensions $30 \times 30$ mm$^2$.  The target was held between the
pole tips of a compact electromagnet that provided a magnetic field
of 0.11 T, sufficient to fully magnetize the Fe layer. To minimize
possible systematic errors, the external magnetic field was
automatically reversed every 600 s.

Table~\ref{paramtable} summarizes the properties of the 2$^+_1$
states, the key aspects of the energy loss of the sulfur beams in
the target, the precession results and the extracted $g$~factors.
The high-$Z$ Au target layer serves to enhance the Coulomb
excitation yield and slow the projectiles to under 800 MeV, while
the thick iron layer results in a long interaction time with the
transient field, maximizing the spin precession. The sulfur
fragments emerge with energies in the range from $\sim 80$~MeV to
$\sim 200$~MeV. Most of this energy spread stems from the energy
width of the radioactive beam.



Projectiles scattering forward out of the target were detected with
a 15.24 cm diameter plastic scintillator phoswich detector placed
79.2 cm downstream of the target position. The maximum scattering
angle of 5.5$^{\circ}$ limits the distance of closest approach to
near the nuclear interaction radius in both the Au and Fe target
layers.
Positioning the particle detector downstream also lowers the
exposure of the $\gamma$-ray detectors to the radioactive decay of
the projectiles.


To detect de-excitation $\gamma$ rays, the target chamber was
surrounded by 14 HPGe detectors of the Segmented Germanium Array
(SeGA) \cite{Muel2001}. The SeGA detectors were positioned with the
crystal centers 24.5 cm from the target position. Six pairs of
detectors were fixed at symmetric angles $(\pm\theta,\phi) =
(29^{\circ}, 90^\circ)$, $(40^{\circ}, 131^\circ)$, $(60^{\circ},
61^\circ)$, $(139^{\circ}, 46^\circ)$, $(147^{\circ}, 143^\circ)$,
and $(151^{\circ}, 90^\circ)$, where $\theta$ is the polar angle
with respect to the beam axis and $\phi$ is the azimuthal angle
measured from the vertical direction, which coincides with the
magnetic field axis. Each $\pm \theta$ pair is in a plane that
passes through the center of the target; $(-\theta,\phi) =
(\theta,\phi + 180^\circ)$. Two more detectors were placed at
$\theta = 90^{\circ}$ and $\theta = 24^{\circ}$. All 14 detectors
were used to measure the $\gamma$-ray angular correlations
concurrently with the precessions. Since the precession angles are
small, the unperturbed angular correlation can be reconstructed by
adding the data for the two directions of the applied magnetic
field.



\begin{figure}[tbp]
\begin{center}
\begin{tabular}{cc}
\resizebox{44mm}{!}{\includegraphics{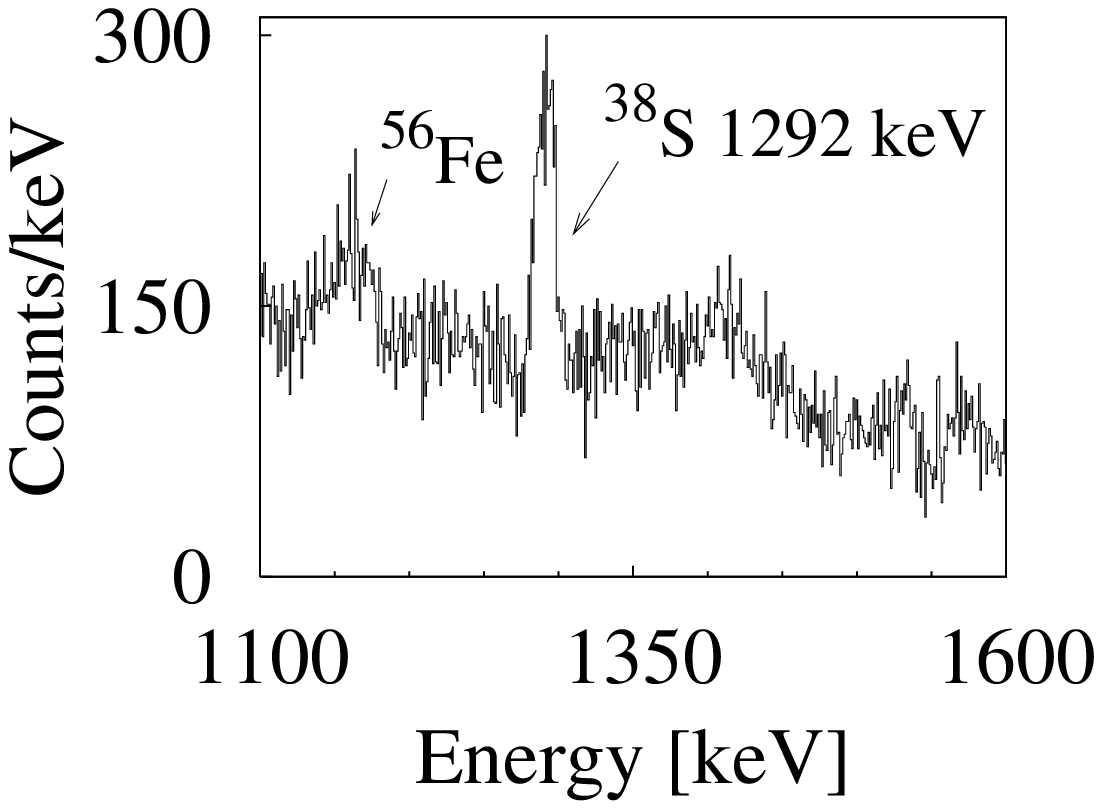}} &
\resizebox{42mm}{!}{\includegraphics{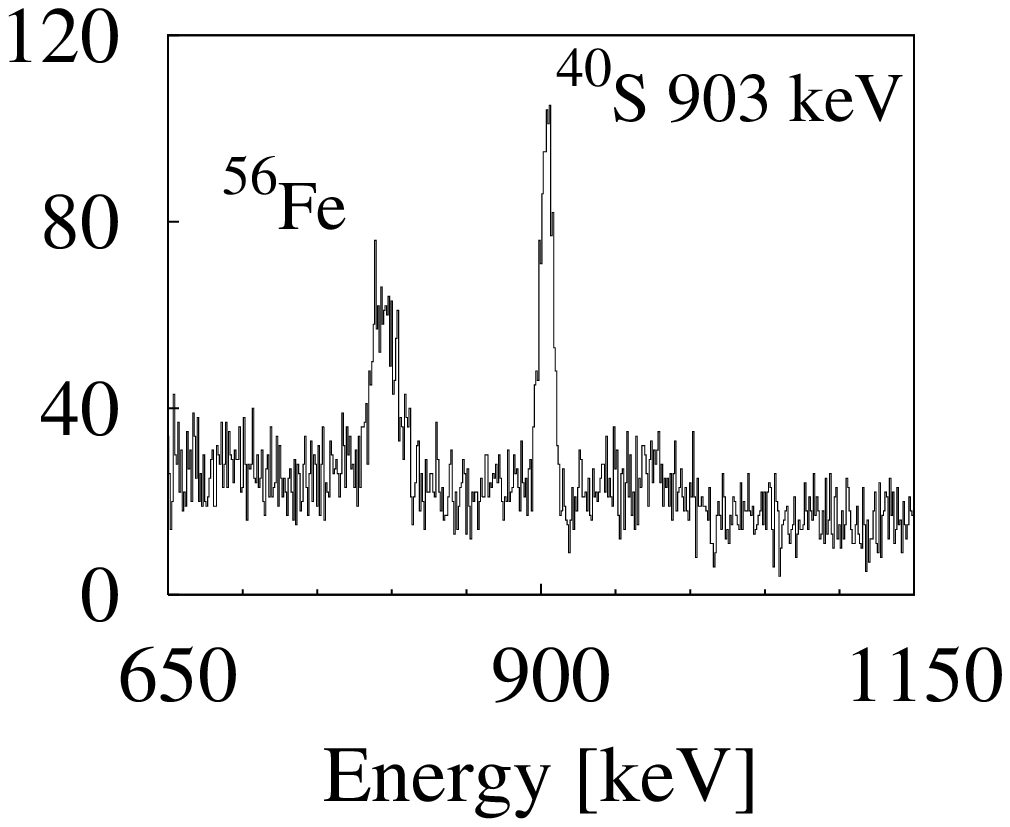}}
\end{tabular}
\caption{Doppler-corrected $\gamma$-ray spectra at
$\theta=40^{\circ}$ for $^{38}$S (left) and $^{40}$S (right). The
sulfur and iron peaks are labeled. The broad feature to the right of
the 1292 keV line is mainly its Doppler tail, due to decays within
the target. } \label{3840Sspectra}
\end{center}
\end{figure}

Coincidences between the phoswich particle detector and SeGA were
recorded, and $\gamma$-ray spectra gated on sulfur recoils and
corrected for random coincidences were produced.  Doppler-corrected
spectra were also produced using the angular information from the
SeGA detector segments and the particle energy information from the
phoswich detector on an event-by-event basis, which is essential
because of the spread in particle velocities.
Figure~\ref{3840Sspectra} shows examples of the $\gamma$-ray
spectra. From the measured Doppler shift of the deexcitation
$\gamma$ rays in the laboratory frame, the average after-target ion
velocities were determined.
The velocity distribution of the exiting $^{40}$S ions
was also measured by shifting the phoswich detector by $\pm 15$~cm
from its normal position and observing the change in the flight
times of the projectiles. These procedures firmly establish that the
sulfur ions were slowed through the peak of the TF strength at
$Zv_0$ into the region where it has been well characterized
\cite{spe02,Stuc2004a}.


The 2$^+$ peak areas averaged 925 counts/detector per field
direction for $^{38}$S and 400 counts/detector per field direction
for $^{40}$S, in each of the six angle pairs of SeGA detectors used
for extracting the precessions.



\begin{figure}[btp]
\begin{center}
\begin{tabular}{cc}
\resizebox{45.5mm}{!}{\includegraphics{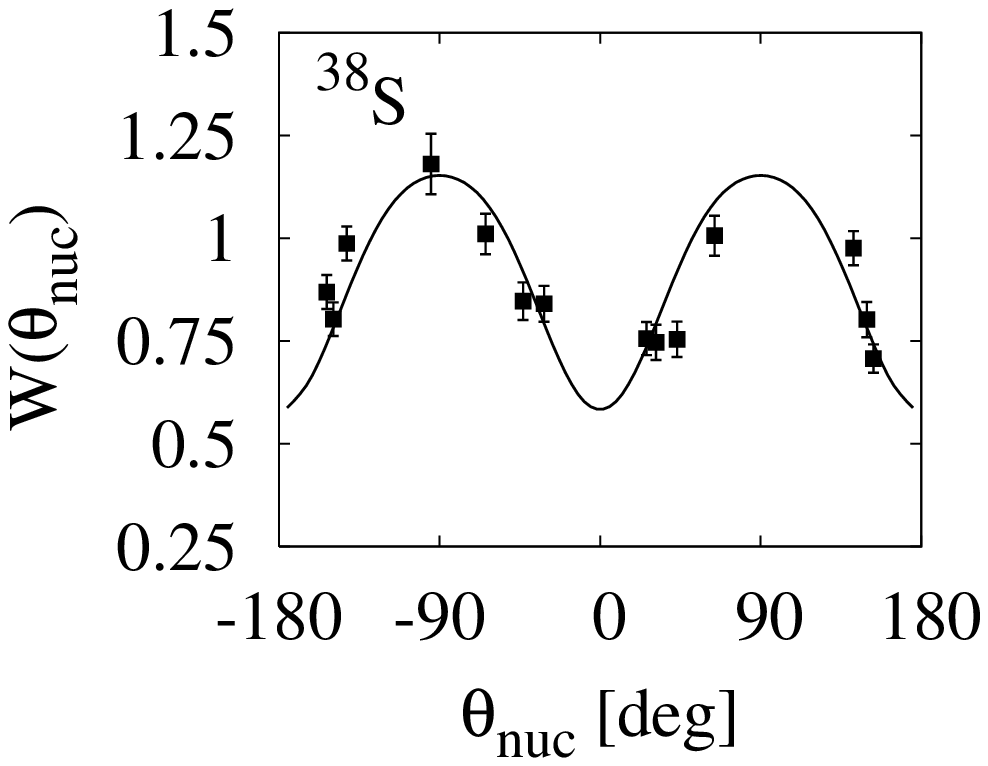}} &
\resizebox{36mm}{!}{\includegraphics{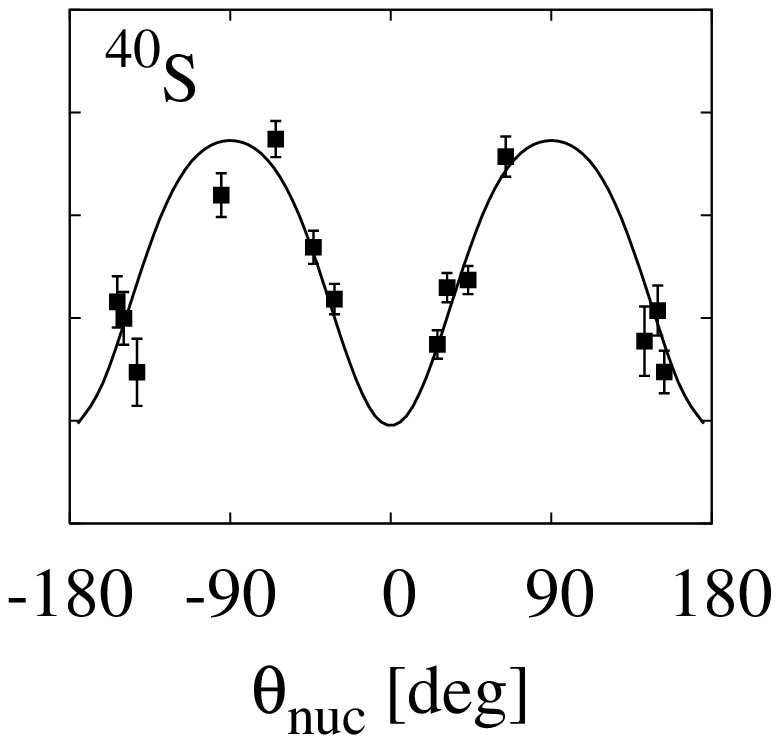}}
\end{tabular}
\caption{Angular correlations for $^{38}$S and $^{40}$S in the frame
of the projectile nucleus. Data are normalized to the calculated
angular correlation. The difference in anisotropy stems from the
alignment produced by Coulomb excitation, which depends on the ratio
of E(2$^+$) to the beam velocity.} \label{3840Sangdist}
\end{center}
\end{figure}

The angular correlation of $\gamma$ rays was calculated with the
program GKINT \cite{Stuc2005b} using the theory of Coulomb
excitation \cite{Bert2003}.
Recoil-in-vacuum effects were evaluated based on measured
charge-state fractions for sulfur ions
emerging from iron foils 
\cite{Stuc2005c}. Good agreement was found between the calculated
$\gamma$-ray angular correlations and the data, as shown in
Fig~\ref{3840Sangdist}.
The experimental nuclear precession angle, $\Delta \theta_{\rm
exp}$, was extracted from $\gamma$-ray count ratios in pairs of
detectors at $(\pm \theta, \phi)$ for both field directions, using
standard analysis methods as described in \cite{Stuc2005a}.



\begin{table*}
\caption{Nuclear parameters and reaction kinematics.
$B(E2)\!\!\uparrow \ = B(E2; 0^+_{\rm gs} \rightarrow 2^+_1)$.
$\langle E_{i,e} \rangle$ and $\langle v_{i,e} \rangle$ are the ion
kinetic energies and velocities at the entrance and exit of the iron
layer. The effective transient-field interaction time $t_{\rm eff}$
and the spin precession per unit $g$ factor, $(\Delta\theta/g)_{\rm
calc}$, are evaluated for ions that decay after leaving the target.
The experimental $g$~factor is given by $g = \Delta \theta_{\rm exp}
/ (\Delta \theta / g)_{\rm calc}$.
}
\begin{ruledtabular}
\begin{tabular}{*{12}c}
Isotope & E(2$^+_1$) & $B(E2)\uparrow$ & $\tau$(2$^+_1$) & $\langle
E_i \rangle$ & $\langle E_e \rangle$ & $\langle v_i/Zv_0 \rangle$ &
$\langle v_e/Zv_0 \rangle$ & $t_{\rm eff}$ &
$(\Delta\theta/g)_{\rm calc}$ & $\Delta \theta_{\rm exp}$ & $g$\\
 & (keV) & (e$^2$fm$^4$) & (ps) & (MeV) & (MeV) & & & (ps) & (mrad) & (mrad)\\
\hline
$^{38}$S & 1292 & 235(30) & 4.9 & 762 & 123 & 1.75 & 0.71 & 2.98 & $-330$ & $-43(15)$ & $+0.13(5)$ \\
$^{40}$S & 904 & 334(36) & 21 & 782 & 145 & 1.73 & 0.75 & 2.99 & $-339$ & $+5(21)$ & $-0.02(6)$\\
\end{tabular}
\end{ruledtabular}
\protect \label{paramtable}
\end{table*}



An evaluation of $\Delta \theta / g = (-\mu_{\rm N}/\hbar) \int
B_{\rm tf} dt$ is required to extract the $g$~factors. Calculations
were performed using the code GKINT to take into account the
incoming and exiting ion velocities, the energy- and angle-dependent
Coulomb excitation cross sections in both target layers, the
excited-state lifetimes, and the parametrization of the TF strength
in Eq.~(\ref{eq:aes-param}). The results and the $g$ factors
extracted are given in Table~\ref{paramtable}.
These $g$~factor results are  not very sensitive to the somewhat
uncertain behavior of the transient field at the highest velocities
because (i) the ions spend least time interacting with the TF at
high velocity and (ii) the TF strength near $2Zv_0$ is very small.
Furthermore, the positive $g$~factor in $^{38}$S and the essentially
null effect for $^{40}$S are both firm observations, independent of
the transient-field strength. The experimental uncertainties
assigned to the $g$~factors are dominated by the statistical errors
in the $\gamma$-ray count ratios, with a small contribution (10\%)
from the angular correlation added in quadrature.

\begin{figure}[btp]
\begin{center}
\resizebox{75mm}{!}{\includegraphics{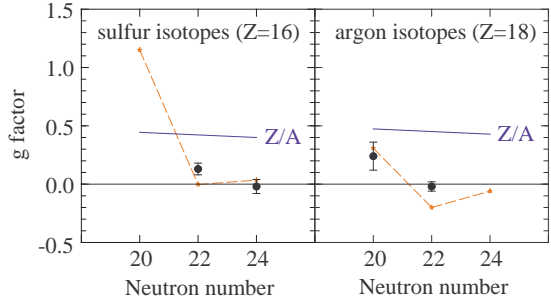}} \caption{Theoretical
$g$~factors compared with experiment. The previous results for
$^{38}$Ar and $^{40}$Ar are from Refs. \protect \cite{Spei2005} and
\protect \cite{Stef2005}, respectively.} \label{theoryfig}
\end{center}
\end{figure}

Shell model calculations were performed for $^{36}_{16}$S$_{20}$,
$^{38}_{16}$S$_{22}$ and $^{40}_{16}$S$_{24}$, and their isotones
$^{38}_{18}$Ar$_{20}$, $^{40}_{18}$Ar$_{22}$ and
$^{42}_{18}$Ar$_{24}$, using the code OXBASH \cite{oxbash} and the
$sd$-$pf$ model space where (for $N\geq20$) valence protons are
restricted to the $sd$ shell and valence neutrons are restricted to
the $pf$ shell. The Hamiltonian was that developed in
Ref.~\cite{Numm2001} for neutron-rich nuclei around $N=28$. These
calculations reproduce the energies of the low-excitation states to
within 200 keV. With standard effective charges of $e_p \sim 1.5$
and $e_n \sim 0.5$ they also reproduce the measured $B(E2)$ values.
The $g$~factors of the 2$^+_1$ states were evaluated using the bare
nucleon $g$~factors. The calculated $g$~factors are compared with
experimental results in Fig.~\ref{theoryfig}. Overall the level of
agreement between theory and experiment is satisfactory given the
extreme sensitivity to configuration mixing and the near cancelation
of proton and neutron contributions in the $N=22,24$ isotones (see
below).

In the $N=20$ isotones, $^{36}$S and $^{38}$Ar, the 2$^+_1$ state is
a pure proton excitation for our model space. Two neutrons have been
added in the $fp$ shell in the $N=22$ isotones $^{38}$S and
$^{40}$Ar. Since $^{36}$S is almost doubly magic, the initial
expectation might be that the first-excited state of $^{38}$S would
be dominated by the neutron $f_{7/2}$ configuration weakly coupled
to the $^{36}$S core, resulting in a $g$~factor near $-0.3$. In
contrast, the near zero theoretical $g$~factor and the small but
positive experimental $g$~factor require additional proton
excitations, which indicates strong coupling between protons and
neutrons - one of the prerequisites for the onset of deformation.
For $N=22,24$ the shell model predicts a cancelation of the proton
and neutron contributions to the moment; in terms of $g^{\rm
th}=g^{\rm th}_{\rm proton} + g^{\rm th}_{\rm neutron}$,
$g(^{38}{\rm S})=-0.003=0.298-0.301$, $g(^{40}{\rm
S})=0.035=0.276-0.241$, $g(^{40}{\rm Ar})=-0.200=0.164-0.364$ and
$g(^{42}{\rm Ar})=-0.060=0.220-0.280$. The proton contributions to
the $g$~factors are dominated by the orbital component but the
substantial neutron contributions originate entirely with the
intrinsic spin associated with a dominant occupation of the neutron
$f_{7/2}$ orbit. Some tuning of the Hamiltonian may be required to
reproduce the sign of the $g$~factor in $^{38}$S, which is very
sensitive to the separation of the proton $s_{1/2}$ and $d_{3/2}$
orbitals, for example. The dependence of the $g(2^+)$ in $^{40}$Ar
on the basis space, the interaction, and the choice of effective
nucleon $g$~factors, has been investigated in Ref.~\cite{Stef2005}.

As noted above, Coulomb-excitation studies and the level scheme of
$^{40}$S suggest that it is deformed. Supporting this
interpretation, the shell model calculations predict consistent
intrinsic quadrupole moments when derived from either the $B(E2)$ or
the quadrupole moment, $Q(2^+_1)$, implying a prolate deformation of
$\beta \approx +0.3$, in agreement with the value deduced from the
experimental $B(E2)$~\cite{Sche1996,Reta1997}. But the near zero
magnetic moment does not conform to the usual collective model
expectation of $g \sim Z/A$. Since the shell model calculations
reproduce both the electric and magnetic properties of the 2$^+_1$
state they give insight into the reasons for this unprecedented
magnetic behavior in an apparently deformed nucleus. The essential
difference between the deformed neutron-rich sulfur isotopes and the
deformed nuclei previously encountered (i.e. either light nuclei
with $N=Z$ or heavier deformed nuclei) is that the spin
contributions to the magnetic moments are relatively more important,
especially for the neutrons. In comparison to $^{40}$S, the 2$^+_1$
state in the $N=20$ nucleus $^{32}_{12}$Mg$_{20}$ has a similar
excitation energy, lifetime and $B(E2)$. However the $g$~factor in
$^{32}$Mg might be closer to that of a conventional collective
nucleus since the $N=20$ shell closure is known to vanish far from
stability.

We thank the NSCL operations staff for providing the primary and
secondary beams for the experiment. This work was supported by NSF
grants PHY-01-10253, PHY-99-83810, and PHY-02-44453. AES, ANW, and
PMD acknowledge travel support from the ANSTO AMRF scheme
(Australia).

\bibliography{AESsulfurHVTF}

\begin{thebibliography}{10}
\expandafter\ifx\csname bibnamefont\endcsname\relax
  \def\bibnamefont#1{#1}\fi
\expandafter\ifx\csname bibfnamefont\endcsname\relax
  \def\bibfnamefont#1{#1}\fi
\expandafter\ifx\csname url\endcsname\relax
  \def\url#1{\texttt{#1}}\fi
\expandafter\ifx\csname urlprefix\endcsname\relax\def\urlprefix{URL }\fi
\providecommand{\eprint}[2][]{\url{#2}}

\bibitem{gre04}
\bibfnamefont{S.}~\bibnamefont{Gr{\'e}vy}, \emph{et~al.}, Phys.\ Lett.\ B
  \textbf{594}, 252 (2004).

\bibitem{fri05}
\bibfnamefont{J.}~\bibnamefont{Fridmann}, \emph{et~al.}, Nature \textbf{435},
  922 (2005).

\bibitem{Sche1996}
\bibfnamefont{H.}~\bibnamefont{{Scheit}}, \emph{et~al.}, \prl \textbf{77}, 3967
  (1996).

\bibitem{Glasm1997}
\bibfnamefont{T.}~\bibnamefont{{Glasmacher}}, \emph{et~al.}, Phys.\ Lett.\ B
  \textbf{395}, 163 (1997).

\bibitem{Wing2001}
\bibfnamefont{J.}~\bibnamefont{Winger},
  \bibfnamefont{P.}~\bibnamefont{Mantica},
  \bibfnamefont{R.}~\bibnamefont{Ronningen}, \bibnamefont{and}
  \bibfnamefont{M.}~\bibnamefont{Caprio}, \prc \textbf{64}, 064318 (2001).

\bibitem{Sohl2002}
\bibfnamefont{D.}~\bibnamefont{{Sohler}}, \emph{et~al.}, \prc \textbf{66},
  054302 (2002).

\bibitem{Cott1998}
\bibfnamefont{P.~D.} \bibnamefont{{Cottle}} \bibnamefont{and}
  \bibfnamefont{K.~W.} \bibnamefont{{Kemper}}, \prc \textbf{58}, 3761 (1998).

\bibitem{rod02}
\bibfnamefont{R.}~\bibnamefont{Rodr{\'i}guez-Guzm{\'a}n},
  \bibfnamefont{J.}~\bibnamefont{Edigo}, \bibnamefont{and}
  \bibfnamefont{L.}~\bibnamefont{Robledo}, \prc \textbf{65}, 024304 (2002).

\bibitem{cau04}
\bibfnamefont{E.}~\bibnamefont{Caurier},
  \bibfnamefont{F.}~\bibnamefont{Nowacki}, \bibnamefont{and}
  \bibfnamefont{A.}~\bibnamefont{Poves}, Nucl. Phys. \textbf{A 742}, 14 (2004).

\bibitem{Stuc2004a}
\bibfnamefont{A.~E.} \bibnamefont{{Stuchbery}}, \prc \textbf{69}, 064311
  (2004).

\bibitem{Glasm1998}
\bibfnamefont{T.}~\bibnamefont{Glasmacher}, Annu.\ Rev.\ Nucl.\ Sci.
  \textbf{48}, 1 (1998).

\bibitem{Kol1980}
\bibfnamefont{N.}~\bibnamefont{{Benczer-Koller}},
  \bibfnamefont{M.}~\bibnamefont{{Hass}}, \bibnamefont{and}
  \bibfnamefont{J.}~\bibnamefont{{Sak}}, Annu.\ Rev.\ Nucl.\ Sci. \textbf{30},
  53 (1980).

\bibitem{spe02}
\bibfnamefont{K.~H.} \bibnamefont{{Speidel}},
  \bibfnamefont{O.}~\bibnamefont{{Kenn}}, \bibnamefont{and}
  \bibfnamefont{F.}~\bibnamefont{{Nowacki}}, Prog. Part. Nucl. Phys.
  \textbf{49}, 91 (2002).

\bibitem{Stuc2005a}
\bibfnamefont{A.~E.} \bibnamefont{{Stuchbery}}, \emph{et~al.}, Phys.\ Lett.\ B
  \textbf{611}, 81 (2005).

\bibitem{Morr2003}
\bibfnamefont{D.~J.} \bibnamefont{{Morrissey}}, \emph{et~al.}, Nucl.\ Inst.\
  Meth.\ Phys.\ Res.\ B \textbf{204}, 90 (2003).

\bibitem{Muel2001}
\bibfnamefont{W.~F.} \bibnamefont{{Mueller}}, \emph{et~al.}, Nucl.\ Inst.\
  Meth.\ Phys.\ Res.\ A \textbf{466}, 492 (2001).

\bibitem{Stuc2005b}
\bibfnamefont{A.~E.} \bibnamefont{{Stuchbery}}, \emph{{Some notes on the
  program GKINT: Transient-field $g$-factor kinematics at intermediate
  energies}}, {Department of Nuclear Physics, The Australian National
  University, report no. ANU-P/1678} (2005).

\bibitem{Bert2003}
\bibfnamefont{C.~A.} \bibnamefont{{Bertulani}}, \bibfnamefont{A.~E.}
  \bibnamefont{{Stuchbery}}, \bibfnamefont{T.~J.} \bibnamefont{{Mertzimekis}},
  \bibnamefont{and} \bibfnamefont{A.~D.} \bibnamefont{{Davies}}, \prc
  \textbf{68}, 044609 (2003).

\bibitem{Stuc2005c}
\bibfnamefont{A.~E.} \bibnamefont{{Stuchbery}}, \bibfnamefont{P.~M.}
  \bibnamefont{{Davidson}}, \bibnamefont{and} \bibfnamefont{A.~N.}
  \bibnamefont{{Wilson}}, Nucl.\ Instr.\ Meth.\ Phys.\ Res.\ B \textbf{243},
  265 (2006).

\bibitem{oxbash}
\bibfnamefont{B.~A.} \bibnamefont{Brown}, \emph{et~al.}, \emph{Oxbash for
  Windows PC}, Michigan State University, report no. MSU-NSCL 1289 (2004).

\bibitem{Numm2001}
\bibfnamefont{S.}~\bibnamefont{Nummela}, \emph{et~al.}, \prc \textbf{63},
  044316 (2001).

\bibitem{Spei2005}
\bibfnamefont{K.-H.} \bibnamefont{Speidel}, \emph{et~al.}, Phys.\ Lett.\ B
  \textbf{632}, 207 (2006).

\bibitem{Stef2005}
\bibfnamefont{E.}~\bibnamefont{Stefanova}, \emph{et~al.}, \prc \textbf{72},
  014309 (2005).

\bibitem{Reta1997}
\bibfnamefont{J.}~\bibnamefont{Retamosa},
  \bibfnamefont{E.}~\bibnamefont{Caurier},
  \bibfnamefont{F.}~\bibnamefont{Nowacki}, \bibnamefont{and}
  \bibfnamefont{A.}~\bibnamefont{Poves}, \prc \textbf{55}, 1266 (1997).

\end{thebibliography}

\end{document}